%% file: main.tex
\documentclass[runningheads]{llncs}
\usepackage[T1]{fontenc}
\usepackage{times}
\usepackage{epsfig}
\usepackage{graphicx}
\usepackage{amsmath}
\usepackage{amssymb}
\usepackage{caption}
\usepackage{cuted}

\usepackage{cite}
\usepackage{amsmath,amssymb,amsfonts}
\usepackage{algorithmic}
\usepackage{graphicx}
\usepackage{textcomp}

\usepackage{booktabs}
\usepackage{caption}
\usepackage{subcaption}

\usepackage{color}
\usepackage{float}
\usepackage[pagebackref=true,breaklinks=true,colorlinks,bookmarks=false]{hyperref}

\usepackage{tabularx,booktabs}
\newcolumntype{Y}{>{\centering\arraybackslash}X X}
\newcommand{\mycomment}[1]{}

\usepackage[font=small,labelfont=bf,tableposition=top]{caption}
\usepackage[capitalize]{cleveref}
\usepackage{multirow}
\usepackage{xcolor}
\renewcommand\fbox{\fcolorbox{white}{white}}
\begin{document}
\title{Adaptive Enhancement of Extreme Low-Light Images}
%
%
\author{Evgeny Hershkovitch Neiterman \thanks{Each of the first two authors contributed equally.} \and
Michael Klyuchka \and
Gil Ben-Artzi}
\authorrunning{E. Hershkovitch Neiterman et al.}
%
\institute{Department of Computer Science,Ariel University, Israel}
\maketitle              
\begin{abstract}
Existing methods for enhancing dark images captured in a very low-light environment assume that the intensity level of the optimal output image is known and already included in the training set. However, this assumption often does not hold, leading to output images that contain visual imperfections such as dark regions or low contrast. To facilitate the training and evaluation of adaptive models that can overcome this limitation, we have created a dataset of 1500 raw images taken in both indoor and outdoor low-light conditions. Based on our dataset, we introduce a deep learning model capable of enhancing input images with a wide range of intensity levels at runtime, including ones that are not seen during training. Our experimental results demonstrate that our proposed dataset combined with our model can consistently and effectively enhance images across a wide range of diverse and challenging scenarios.
\end{abstract}
\section{Introduction}
\label{sec:intro}

Images captured in low light are characterized by low photon counts, which results in a low signal-to-noise ratio (SNR). Setting the exposure level while capturing an image can be done by the user in manual mode, or automatically by the camera in auto exposure (AE) mode. In manual mode, the user can adjust the ISO, f-number, and exposure time. In auto exposure (AE) mode, the camera measures the incoming light based on through-the-lens (TTL) metering and adjusts the exposure values (EVs), which refers to configurations of the above parameters.  

We consider the problem of enhancing a dark image captured in an extremely low-light environment, based on a single image~\cite{chen2018learning}. In a dark environment, adjusting the parameters to increase the SNR has its own limitations. For example, high ISO increases the noise as well, and lengthening the exposure time might introduce blur. Various approaches have been proposed as post-processing enhancements in low-light image processing ~\cite{yuan2012automatic,hwang2012context,celik2011contextual,lee2013contrast,zhang2012enhancement,hu2014deblurring}. In extreme low light conditions, such methods often fail to produce satisfactory results.

Recent works \cite{remez2017deep,ignatov2017dslr,chen2018learning,wang2019underexposed,xu2020learning}  introduce data-driven approaches to replace the traditional image signal processing pipeline and learn a direct mapping from low-exposure input images to well-lit output images. Such models are trained using a fixed intensity level for the output image. Given a dark image, they first multiply its intensity values by a constant factor to increase its brightness and then apply the enhancement model that is trained specifically for this fixed intensity level to produce high-quality image. However, during runtime, it is common for the optimal intensity level of the output image to differ from the trained one and the model outputs less visually appealing images.

\begin{figure}[t]
\begin{center}
   \includegraphics[width=0.24\linewidth]{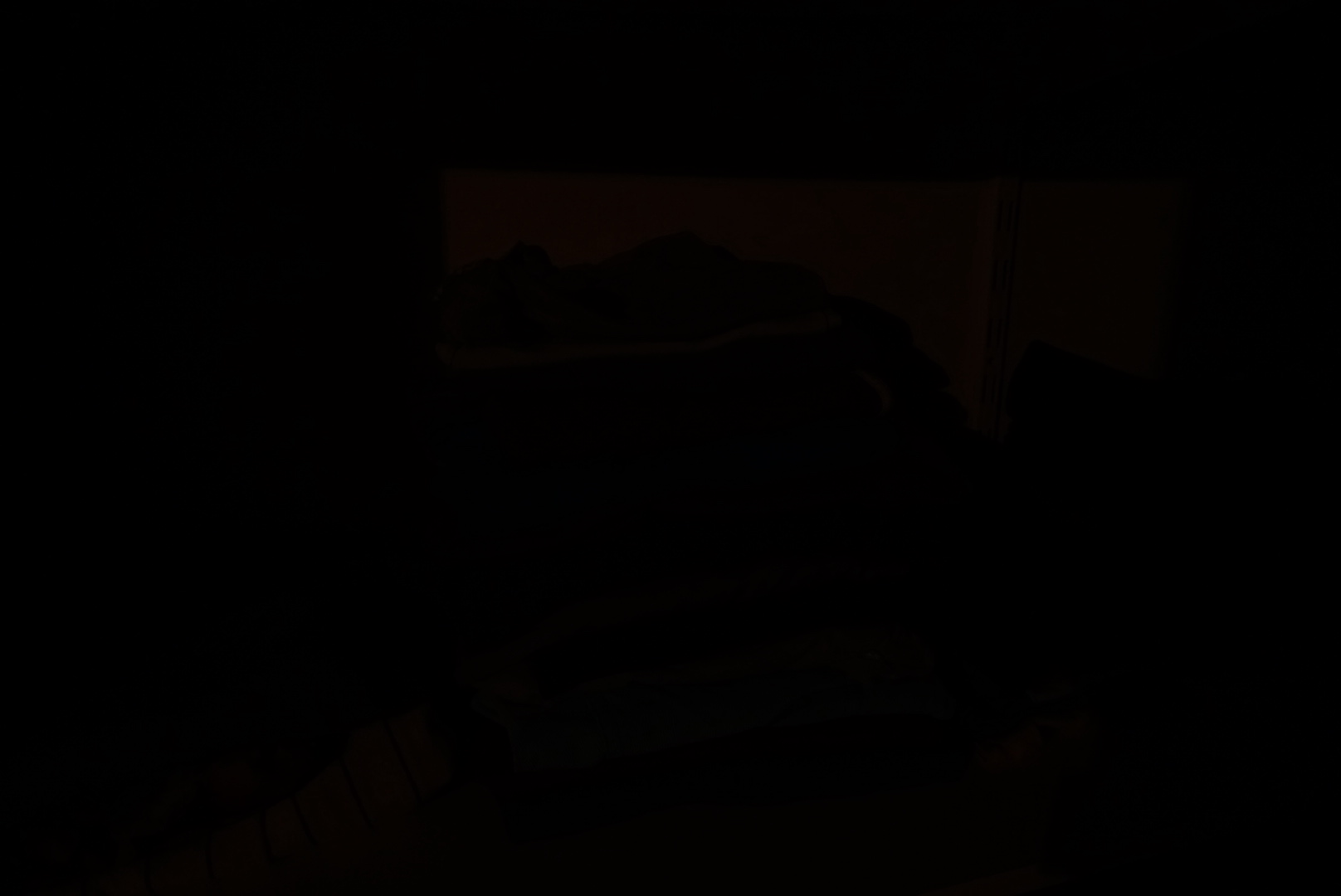}
   \includegraphics[width=0.24\linewidth]{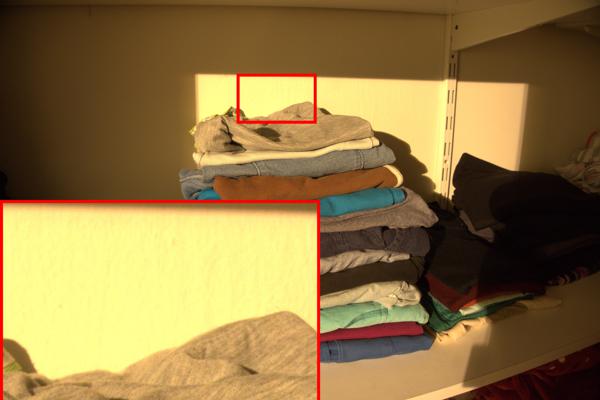}
      \includegraphics[width=0.24\linewidth]{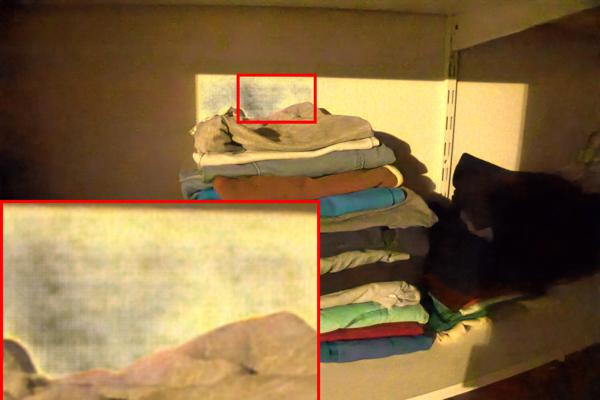}
         \includegraphics[width=0.24\linewidth]{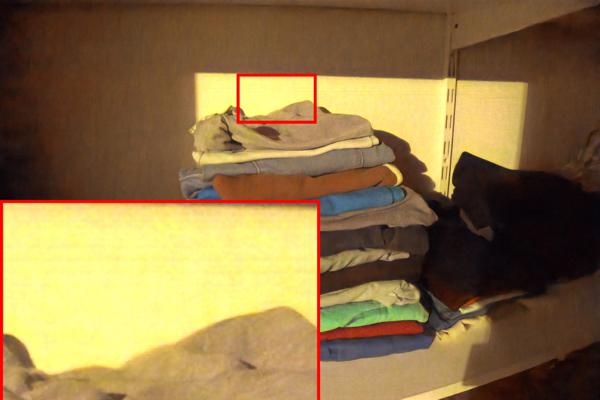}
\end{center}
   \caption{At runtime, the intensity level for optimal restoration of a given dark image might be different from the trained one and can lead to dark areas or low contrast. For a wide range of output image's intensity levels, our model optimizes the enhancement of the input image. (a) left: The input, (b) center left: Ground truth, (c) center right: SID \cite{chen2018learning} enhances the image to a fixed intensity level, which is not optimal for the input image, and as a result, there are noticeable artifacts. (d) right: Our approach adapts the enhancement operation to optimally match any selected intensity levels, thereby reducing the presence of artifacts.}
\label{fig:1}
\end{figure}

We address this limitation by proposing a model that can enhance dark images across a wide range of intensity levels, including those that were not seen during training. Our model achieves this by adaptively adjusting the enhancement operation during runtime to optimally match the selected intensity level from a given range, without requiring retraining of the model. As a result, our model can significantly reduce artifacts in the output image even for previously unseen intensity levels (~\cref{fig:1}).

To enhance real-world images captured in low light, training based on only synthetic noise samples is insufficient. We have collected 1500 raw images captured with five different exposure levels in extreme low-light conditions, in both indoor and outdoor environments, and under various camera parameters. We use fixed exposure times as the exact output intensity of each enhanced image might be different from each other. Each exposure time corresponds to a distinct intensity level, and the range of exposures yields a range of intensity levels. Using the dataset, we show how to train our model such that it can successfully enhance images with different intensity levels at runtime.


Previous adaptive approaches \cite{he2019modulating,he2020interactive} only considered signal-independent noise using sRGB synthetic noise samples. Here we propose an adaptive model with two input parameters to control the output intensity and address both signal-dependent and signal-independent noises. The first parameter controls the intensity level of the output image by simple multiplication. This results in an inevitable amplification of the noises and other artifacts as well. The second parameter adjusts the operation of the image signal processing (ISP) unit to enhance the degradations that are the result of the increase in the intensity, conditioned on the intensity level.


{\bf Contribution.} In pursuit of advancing research in the field and facilitating the development of adaptive models, we have curated a dataset containing 1,500 raw images captured in extremely low-light conditions, comprising indoor and outdoor scenes with diverse exposure levels. We propose and train a model that can produce compelling results for restoring dark images with a wide range of optimal intensity levels, including ones that were not available during training.  Our experimental results, which incorporate both qualitative and quantitative measures, demonstrate that our model along with our dataset improves the enhancement quality of dark images.

\section{Related Work}
\label{sec:works}

\noindent
{\bf Datasets}. A key contribution of our work is a dataset of real-world images that enable training and evaluating multi-exposore models in exterme low light. Unlike existing datasets, we introduce a long-exposure reference image with multiple shorter exposure times for each scene, in both indoor and outdoor scenes, and directly provide the raw sensor data. Our dataset fills the gap and allows the training of an adaptive model in extreme low-light conditions by combining multiple exposures. Our dataset vs. other datasets is compared in \ref{tab:dataset_comp}.


\input{dataset_comp}

\noindent
{\bf Adaptive Restoration Networks}. Adaptive restoration networks can broadly be categorized as models that allow tuning different objectives at runtime \cite{he2019multi} or different restoration levels of the same objective. Dynamic-Net \cite{shoshan2019dynamic} adds specialized blocks directly after convolution layers, which are optimized during the training for an additional objective. CFSNet \cite{wang2019cfsnet} uses branches, each one targeted for a different objective. AdaFM \cite{he2019modulating} adds modulation filters after each convolution layer. Deep Network Interpolation (DNI) \cite{wang2019deep} trains the same network architecture on different objectives and interpolates all parameters. These methods are optimized for well-lit images and as we demonstrate in the experiments, struggle to enhance images captured in extreme low light conditions.


\noindent
{\bf Low-light Image Enhancement}. Widely used enhancement methods are histogram equalization, which globally balances the histogram of the image; and gamma correction, which increases the brightness of dark pixels. More advanced methods include illumination map estimation \cite{guo2016lime},  semantic map enhancement \cite{yan2016automatic}, bilateral learning \cite{gharbi2017deep}, multi-exposure \cite{cai2018learning,ying2017bio,afifi2021learning}, Retinex model \cite{wei2018deep,zhang2019kindling,fu2016weighted,cai2017joint} and unpaired enhancement \cite{jiang2021enlightengan}. In contrast to these methods, we consider an extreme low-light environment with very low SNR, where the scene is barely visible to the human eye. Chen \cite{chen2018learning}  has introduced an approach to extreme low-light imaging by replacing the traditional image processing pipeline with a deep learning model based on raw sensor data. Wang \cite{wang2019underexposed} introduced a neural network for enhancing underexposed photos by incorporating an illumination map into their model, while Xu \cite{xu2020learning} presented a model for low-light image enhancement based on frequency-based decomposition.  These methods are optimized to output an enhanced image with a fixed exposure. In cases where the user requires a change in the exposure (intensity level) of the output image, these methods require retraining the models, typically on additional sets of images. In contrast, we introduce an approach that enables continuous setting of the desired exposure at inference time.

\section{Our Approach}

\begin{figure}[t] 
\begin{center}
   \includegraphics[width=1\linewidth]{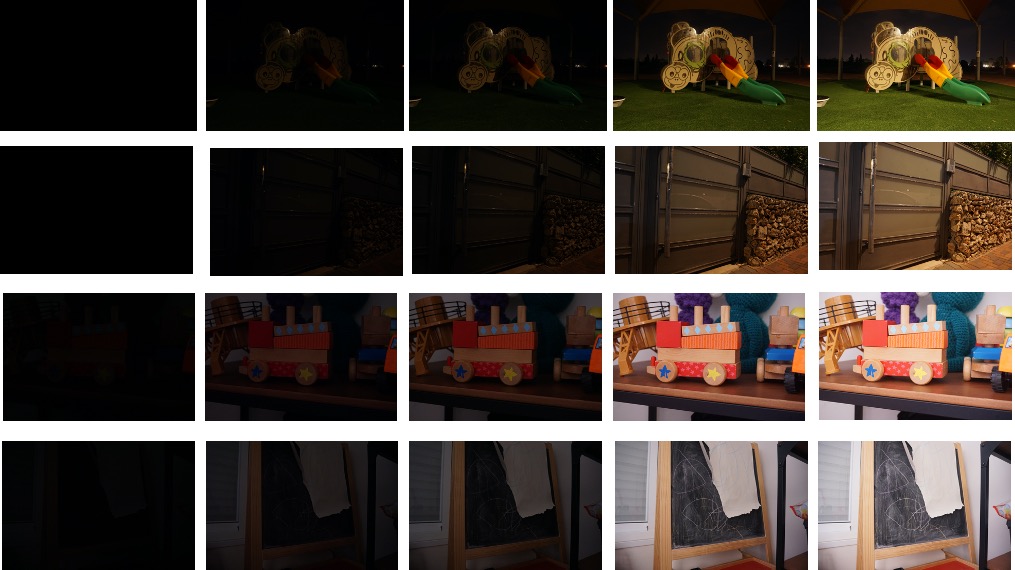}
\end{center}
   \caption{The multi-exposure dataset. The top two rows are images of outdoor scenes, and the bottom two rows are images of indoor scenes. From left to right are exposure times of 0.1s, 0.5s, 1s, 5s, and 10s.}
\label{fig:long}
\label{fig:onecol}
\end{figure}

\begin{figure}[t]
\fbox {
\begin{minipage}{.39\linewidth}
\centering
\includegraphics[width=1\linewidth]{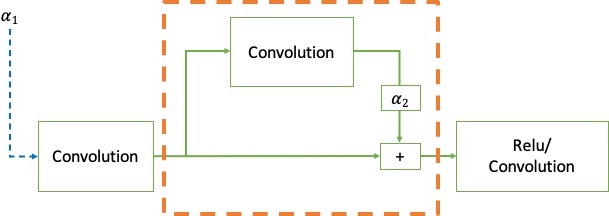}
   \caption{The dashed red rectangle is the modulation module. The enhancement parameter $\alpha_2$ represents a weighted sum between the feature map of the initial and final exposure levels. The blue dashed line is to emphasize that the operation of the modulation module is also affected by the $\alpha_1$ parameters which control the brightness of the image.}
   \label{fig:module}
   \end{minipage}}
\fbox {
\begin{minipage}{.59\linewidth}
\centering
    \includegraphics[width=1\linewidth]{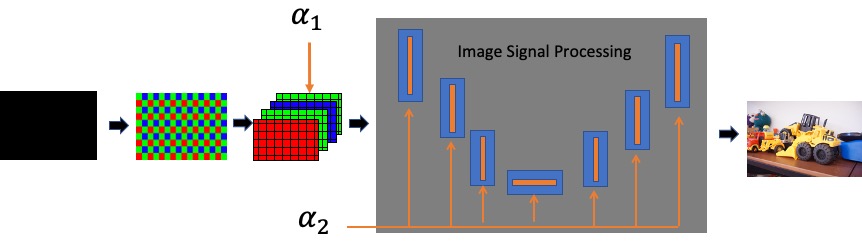}
\caption{The architecture of our network. There are two input parameters, $\alpha_1$ (brightness) and $\alpha_2$ (enhancement). $\alpha_1$ controls the brightness of the raw input data. $\alpha_2$ modulates the weights of the filters and tunes the network, which operates as an Image Signal Processing (ISP) unit. We train the model for an initial and final exposure level, where for each value of $\alpha_1$ there is a single value of $\alpha_2$. At inference time, each parameter can be set independently of the other.} 
\label{fig:arch}
\end{minipage}}

\end{figure}

\subsection{Multi-Exposure Extreme Low-Light Dataset (ME2L)}
\label{sec:our_dataset}

We collected a total of 1500 images. In order to capture a variety of realistic low-light conditions and cover a broad range of scenes with extreme low-light conditions, the images were captured in both indoor and outdoor scenes. The images were captured over different days in multiple locations. We captured five different exposures for each of the scenes - 0.1s, 0.5s, 1s, 5s, and 10s resulting in a range of intensity levels. Various scenes have different intensity levels. By training our model on all the images it learns how to optimize the whole range of intensity levels. 

The outdoor images were captured late at night under moonlight or street lighting. The indoor images were captured in closed rooms with indirect illumination. Generally, the lowest exposure image in both indoor and outdoor scenes is completely dark and no details of the scene can be observed.   

All the scenes in the dataset are static to accommodate the long exposure. For each scene, similar to \cite{chen2018learning}, the settings of the camera were adjusted to optimize the longest-exposure image. We used a tripod and a mirrorless camera to capture the exact same scene without any misalignment. At each scene, after the long exposure image was optimally captured, we used a smartphone application to decrease the exposure and capture the images without touching the camera or changing the camera's parameters. After capturing the images, we manually verified that the images are aligned and the long-exposure reference images are of high perceptual quality.  

The images were captured using a Sony $\alpha$5100 with a Bayer sensor and a resolution of 6000 $\times$ 4000. ~\Cref{fig:long} shows samples from our dataset.

\subsection{Our Model}

The goal of existing extreme low light approaches is to find a function that can map at inference time a data point from raw data space to a single data point in the sRGB space, denoted as $f: \mathcal{Y} \rightarrow \mathcal{Y}_{rgb}$, where $\mathcal{Y}_{rgb}$ is the sRGB space. This approach leads to inaccurate results in cases where the optimal intensity level of the output image is not the same as in the training and may result in noticeable artifacts. Direct change of the signal's mean by a multiplication and applying the same model does not result in the desired outcome, since the noise in raw sensor data results from two main sources: signal-dependent noise and signal-independent. The first one is referred to as shot noise, which is related to the uncertainty that is a property of the underlying signal itself, describing the photon arrival statistics. The second one is read-noise, which is the result of uncertainty generated by the electronics of the camera when the charge stored is read out. The shot noise is a Poisson random variable, whose mean is the expected number of photons per unit time interval, describing the true light intensity. The read noise is a Gaussian random variable with zero mean whose variance is fixed.  

The heteroscedastic Gaussian model is a more widely acknowledged alternative to the Poisson-Gaussian model, which substitutes the Poisson component with a Gaussian distribution whose variance is signal-dependent:

\begin{equation}
    y \sim \mathcal{N}(x, \beta_{read}+\beta_{shot}x), 
    \label{eq:heter1}
\end{equation}

where $y  \in \mathcal{Y}$ is the observed (raw) intensity at a pixel in the raw data space $\mathcal{Y}$, $x$ is the original (unknown) signal, $\beta_{shot}$ is proportional to the analog gain ($g_a$) and digital gain ($g_d$) and $\beta_{read}$ is proportional to the sensor readout variance ($\sigma_r^2$) and digital gain: $\beta_{read}=g_d^2 \sigma_r^2, ~~~~~~ \beta_{shot}=g_d g_a$.



It is therefore evident from \cref{eq:heter1} that unlike previous methods, adding a single noise source (e.g. Gaussian) or using a simple multiplication to adjust the image intensity is not equivalent to acquiring an image with such original intensity. We propose an alternative approach to enhance both read and shot noises by employing two input parameters each contributing differently to ~\cref{eq:heter1}, a modulation layer~\cite{he2019modulating}, and mapping of a \emph{single} data point from raw data space to \emph{multiple} points in sRGB, each with a different output intensity level.

Our Raw-to-sRGB pipeline is formulated as a function $f: \mathcal{Y} \times \mathcal{R} \times \mathcal{R} \rightarrow \mathcal{Y}_{rgb}$, $y_{rgb} = f (y, \alpha_1, \alpha_2; \theta)$, where $\alpha_1$ is a scalar that sets the mean of the signal in~\cref{eq:heter1} to the desired level by multiplication of the raw data, $\alpha_2$ controls the enhancement level of the Raw-to-sRGB pipeline, $\theta$ represents the parameters of $f(\cdot)$ and $y_{rgb} \in \mathcal{Y}_{rgb}$ is the signal of the sRGB image. The function $f$ is realized by a deep network with modulation layers. To obtain $\theta$, we train our network in two steps. First, the base model is trained to fit the enhanced image with an initial intensity level, without any additional modifications to the existing architecture. Then we freeze the weights of the base model, and each modulation layer ($g$) is inserted after each existing convolutional kernel $g(w,b) \circ X$, where $X$ is the output feature map of existing convolutional kernels in the base network and $w,b$ are weights and bias of the modulation layer's convolutional filter kernel. The network is then fine-tuned to fit the enhanced image with a final intensity level by learning the weights of the additional convolutional kernels. Thus, in our formulation, $\theta$ includes the parameters of both the base network and the modulation layers. During runtime, assuming $w_1$ is the base convolution kernel, $w$ and $b$ are the weights of filter and bias in each modulation layer, the output of the modulation layer is:

\begin{equation}
w_1 + \alpha_2w_1 * w +\alpha_2 b,
\label{eq:alpha_2}
\end{equation}
for the given scalar $0 \leq \alpha_2 \leq 1$ representing the enhancement parameter (\cref{fig:module}). 

To control both noise sources, we set $\alpha_2 \in [0,1]$ such that it linearly corresponds to $\alpha_1$ and $\alpha_2=1$ corresponds to the maximum value of $\alpha_1$. Our key intuition is that for $\alpha_1,\alpha_2 \rightarrow 0$, it is the trained base network (before fine tuning) that produces the most significant output, and it enhances the read noise (\cref{eq:alpha_2,eq:heter1}). During training, both parameters are adjusted according to the ground-truth image. The input arrays' values are multiplied by the $\alpha_1$ parameter, which represents the ratio between the input image's exposure time and the required output image's exposure time, effectively setting the intensity and noise levels of the output. The overall architecture of our network is presented in~\cref{fig:arch}.



Unlike existing adaptive method, we do not operate in sRGB domain for noisy images as it limits the representation power of the architecture~\cite{SIDD_2018_CVPR}. Instead, we operate in the raw domain and employ a U-Net \cite{ronneberger2015u} as our base architecture ($f$). It replaces the entire image signal processing (ISP) pipeline. The input is a short exposure raw image from Bayer sensor data and the output is an sRGB image. The raw Bayer sensor data is packed into four channels, the spatial resolution is reduced by a factor of two in each dimension; and the black level is subtracted. The output is a 12-channel image processed to recover the original resolution of the input image. 

For testing, we set the intensity level ($\alpha_1$) and the enhancement ($\alpha_2$) parameters of the network to the desired exposure and ISP configuration. The input image is multiplied according to the intensity level parameter, resulting in a noisy, brighter image. The weights of the filter and bias in the modulation module after the fine-tuning phase are adjusted according to the value of the enhancement parameter. 

We train the model using L1 loss and the Adam optimizer. The inputs are random 512×512 patches with standard augmentation. The learning rate is $10^{-4}$ for 1000 epochs and then $10^{-5}$ for an additional 1000 epochs, a total of 2000 epochs for the training phase. Fine-tuning the model for the final exposure level requires an additional 1000 epochs.

\section{Experiments}
\noindent
{\bf Baselines}. We compare our results with state-of-the-art adaptive methods \cite{he2019modulating,he2020interactive}. Using our dataset, we train them in accordance with their authors' instructions. The inputs of the compared models were modified to operate on raw images in order to ensure fair comparisons.
The {\bf SID}~\cite{chen2018learning} is the baseline model for extreme low light enhancement, and it enhances dark images to a fixed intensity level.

\noindent
{\bf Evaluation Metrics}. We use 70\%, 10\%, and 20\% of the images for training, validation, and testing, respectively, with uniform sampling and equal representation for indoor and outdoor scenes in each set. The ground truth images are the corresponding long-exposure images processed by LibRaw\footnote{www.libraw.org} to sRGB format.

\input{results_table}

\subsection{Quantitative Comparisons}

\label{QC_explained}

\Cref{results_output_A,results_output_B,results_output_C,results_output_D} presents the PSNR and SSIM metrics for various experiments designed to evaluate the different approaches. Each section (A-D) represents a different experiment. The left column shows the different methods and their training protocols. The input for both training and testing is a dark image with an exposure time of 0.1s. For each method, the ground truth exposure times that were used for training (1s/5s/10s) are shown with each model (using the $\Rightarrow$). 

In ~\cref{results_output_A}.A we train the SID model for every single input and output intensity (and exposure) independently. Note that SID is optimized for a single output only. By testing the model on the same exposure as trained, we obtain the optimal achievable restoration accuracy as the model is specialized on a single intensity level. Testing on other exposures (e.g., training on 5s and testing on ground truth image of 10s by setting $\alpha_1$ to the optimal value) shows that the resulting enhanced image quality is significantly reduced, which is the key limitation of single-output methods. The goal of our approach is to overcome this and achieve high restoration quality over the continuous range of possible exposure times with a single model.

~\Cref{results_output_B}.B evaluates the ability to train single-output approach(\cite{chen2018learning}) to generalize to multiple output exposures. We train the model based on all possible output exposures and evaluate its ability to enhance specific exposure times within the trainable range. It can be seen that using multiple ground-truth exposures with a model that is designed to output only a single one reduces the restoration quality for all the possible outputs.

We compare our approach with state-of-the-art adaptive methods. ~\Cref{results_output_C}.C presents the results for one of the most common use-cases: where the optimal exposure time is within the trainable range in runtime. We train the models to enhance input images with an exposure of 0.1s and a ground truth exposure range of [1s,10s]. At inference time, the models can enhance an image to a range of exposures and the specific one is selected. We evaluate the models with input images of 0.1s and optimal output exposure of 5s. It can be seen that our approach outperforms all other methods.

In real-world scenarios, the actual optimal exposure time of the enhanced image can be outside the trained range. We experiment with such cases, training the models for optimal exposure times of [1s,5s], and testing with input images such that the ground truth exposure time is 10s. The results are presented in ~\cref{results_output_D}.D. As before, our approach achieves the best restoration accuracy. 

\begin{figure*}[t]
    \centering
    \begin{tabular}{cccc}
    Brightness Only & SID \cite{chen2018learning} & Ours & Ground truth \\
    \includegraphics[width=0.24\linewidth]{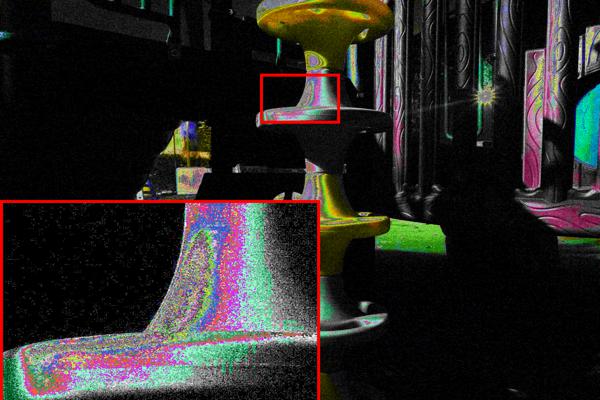} & \includegraphics[width=0.24\linewidth]{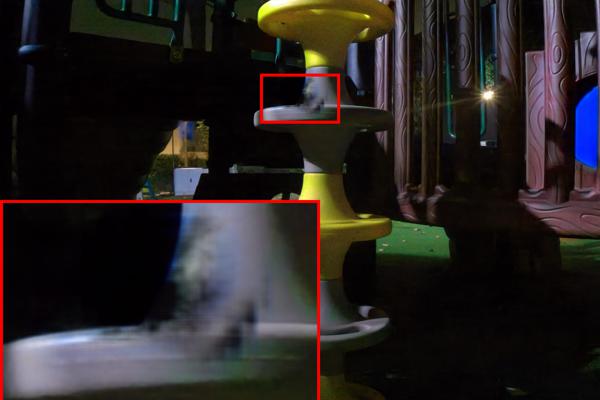}& \includegraphics[width=0.24\linewidth]{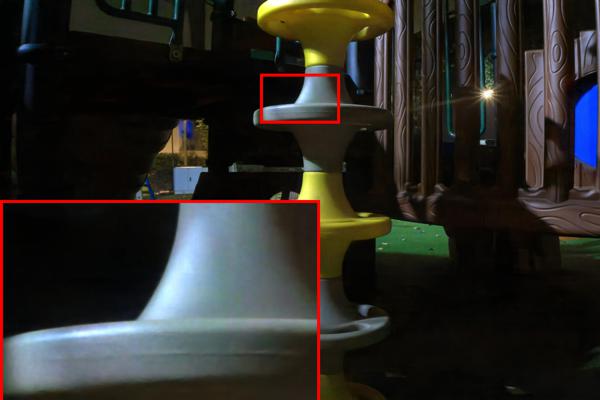}&
    \includegraphics[width=0.24\linewidth]{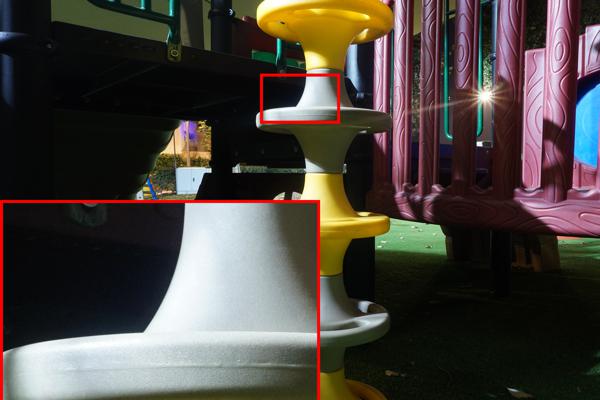}\\
    
    \includegraphics[width=0.24\linewidth]{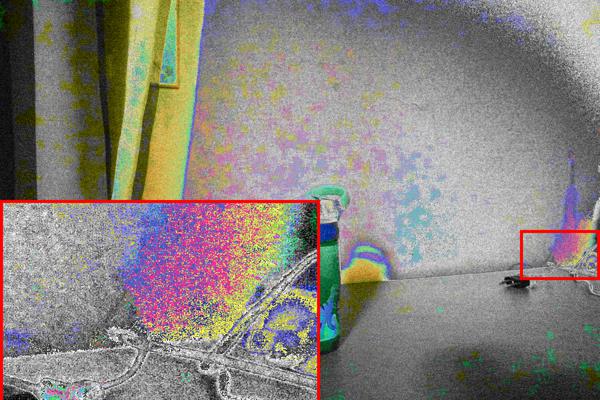} & \includegraphics[width=0.24\linewidth]{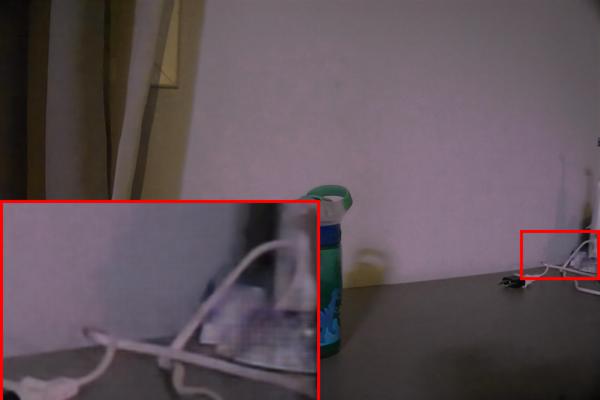}& \includegraphics[width=0.24\linewidth]{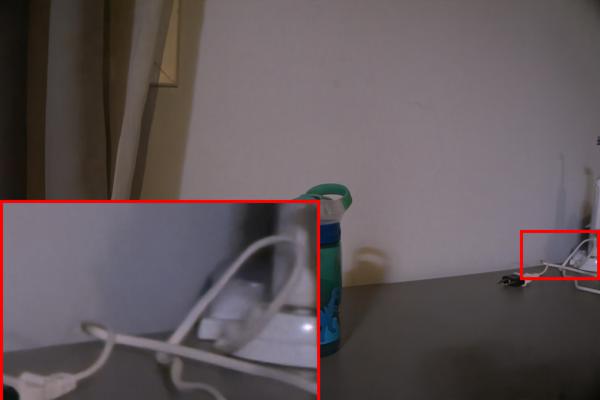}&
    \includegraphics[width=0.24\linewidth]{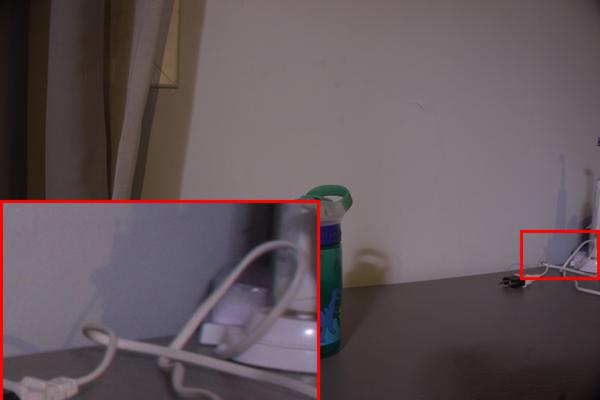}\\
    
    \includegraphics[width=0.24\linewidth]{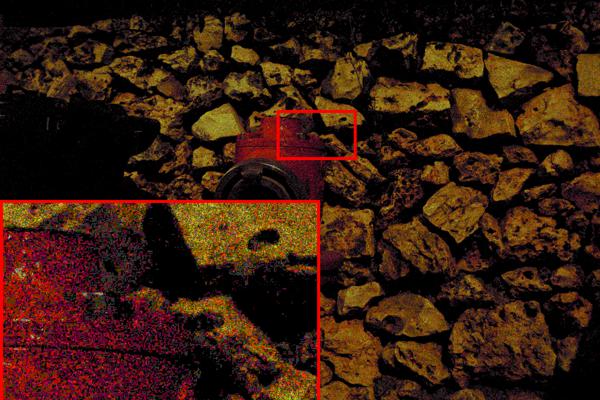} & \includegraphics[width=0.24\linewidth]{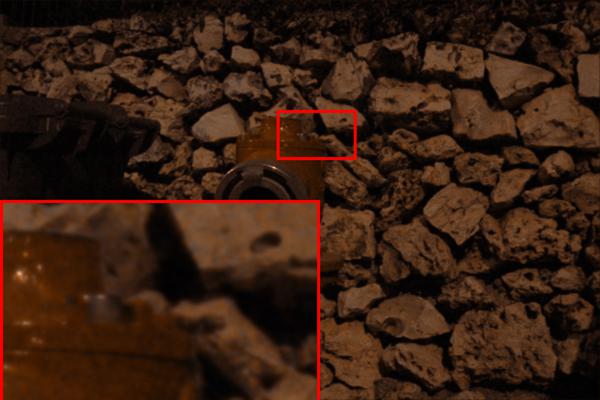}& \includegraphics[width=0.24\linewidth]{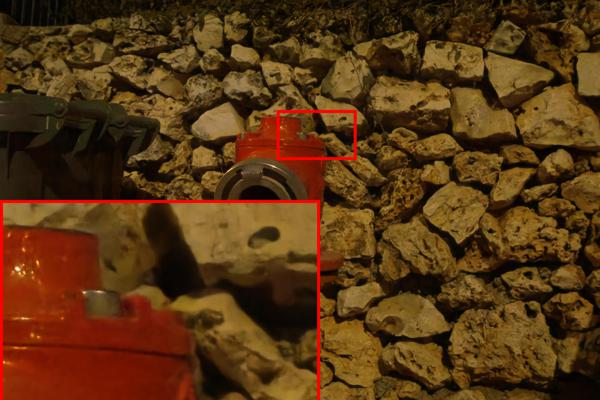}&
    \includegraphics[width=0.24\linewidth]{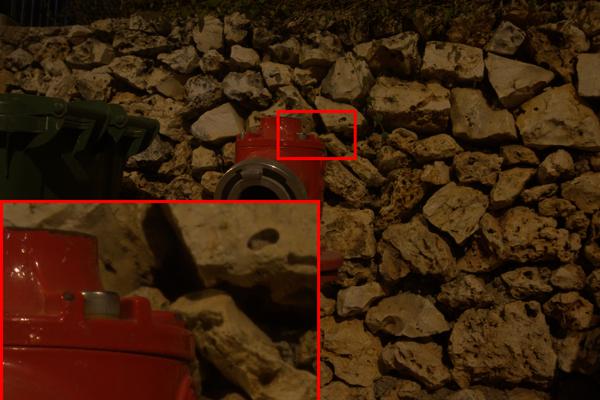}\\
    
    \includegraphics[width=0.24\linewidth]{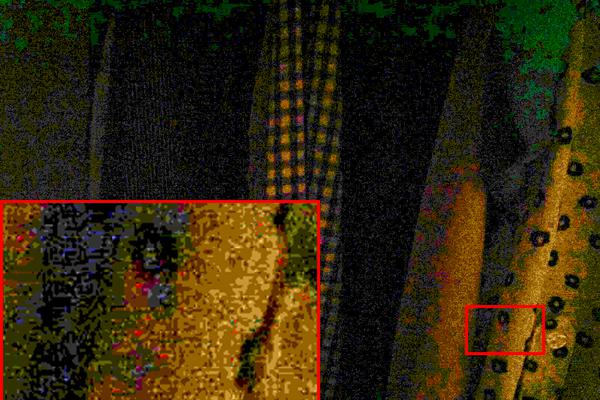} & \includegraphics[width=0.24\linewidth]{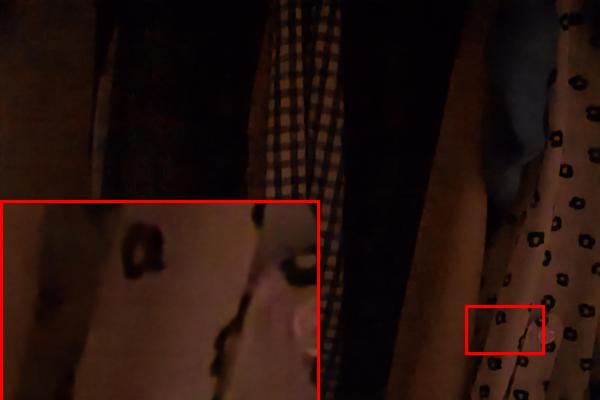}& \includegraphics[width=0.24\linewidth]{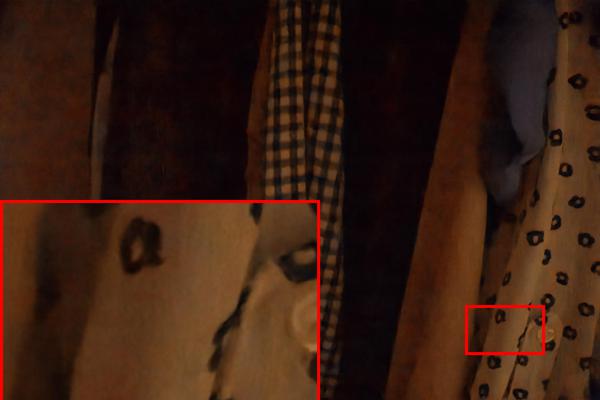}&
    \includegraphics[width=0.24\linewidth]{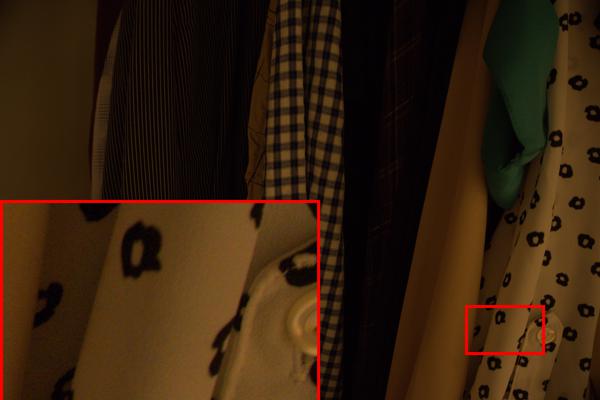}\\

    \end{tabular}
    \caption{The restoration effect of enhancing images to exposure level within the trained range. The first column is obtained by directly adjusting the brightness level to the optimal exposure by multiplication.}
\label{fig:tuning}
\end{figure*}

\subsection{Qualitative Comparisons}
Fig.~\ref{fig:tuning} shows the effect of adjusting the exposure time for a value within the trained range, 5s. The model is trained using input images with an exposure time of 0.1s and ground truth images with exposure times of 1s and 10s. SID was trained on all possible output exposure times. The enhanced images after adjusting the brightness and enhancement parameters are shown. The left column shows the effect of multiplying the intensity of the input images by 50, which is the ratio between the ground truth exposure of the input (0.1s) and the ground truth output (5s). It can be seen that our model successfully removes the artifacts presented by the other approaches.

\subsection{Ablation Study}

{\bf Filter Size.}  We evaluate the sizes of different filters in the modulation module. We consider filter sizes of – 1×1, 3×3, 5×5, and 7×7. We train our base model with an exposure of 0.1s and an output of 1s, then fine-tune it to an output of 10s. The test images are with an exposure of 5s.  

 Table~\ref{filters} shows our comparisons. It can be seen that the most significant gain is achieved when using a filter size of $3\times3$. \\  

\begin{table}
\begin{center}
\begin{tabular}{c|c|c|c}
\hline
   1 $\times$ 1 & 3 $\times$ 3 & 5 $\times$ 5 & 7 $\times$ 7  \\
  \hline
 \hline
  31.87 &32.35  & 32.39 & 32.48  \\
\hline
\end{tabular}
\end{center}
\caption{Filter size comparisons. The model is trained from 0.1s to 1s and fine-tuned to 10s, and tested for an unseen exposure level of 5s.}
\label{filters}
\end{table}

{\bf Tuning Direction.} We evaluate the optimal direction for the tuning. We compare two models. The first one is trained from 0.1s to 1s and fine-tuned for 10s. The second one is trained from 0.1s to 10s and fine-tuned for 1s. 
We compare the results with respect to unseen output images with an exposure time of 5s.
The forward direction from 0.1s to 10s achieved better results than the backward one, with a PSNR of 32.35 vs. 28.2. \\

\section{Conclusion}

Extreme low-light imaging is challenging and has recently gained growing interest. Current methods allow enhancement of dark images, assuming the input exposure and the optimal output exposure are known at inference time, which prevents its adaptation in practical scenarios. We collected a dataset of 1500 images with multiple exposure levels for extreme low-light imaging.  We present an approach that enables continuously controlling of the optimal output exposure levels of the images at runtime, without the need to retrain the model and showed that our model presents promising results on a wide range of both indoor and outdoor images. We believe that our dataset as well as our model will support further research in the field of extreme low-light imaging, making a step forward towards its widespread adoption. 

\bibliographystyle{main}
\bibliography{main}
\end{document}

%% file: dataset_comp.tex
\begin{table}[]
\centering
\resizebox{\textwidth}{!}{%
\begin{tabular}{lllccc}
\hline
\multicolumn{1}{|l|}{Dataset} &
  \multicolumn{1}{l|}{Format} &
  \multicolumn{1}{l|}{\# Images} &
  \multicolumn{1}{l|}{Publicly Available} &
  \multicolumn{1}{l|}{Multi Exposure} &
  \multicolumn{1}{l|}{Extreme Low Light} \\ \hline
DND         \cite{plotz2017benchmarking }   & RAW & 100   & yes & no & no  \\
SIDD        \cite{SIDD_2018_CVPR}           & RAW & 30000 & yes & yes & no \\
LLNet       \cite{lore2016llnet}            & RGB & 169   & yes & no & no  \\
MSR-Net     \cite{shen2017msrnetlowlight}   & RGB & 10000 & no  & no & no  \\
SID        \cite{chen2018learning}         & RAW & 5094  & yes & no & yes \\
SICE        \cite{Cai2018deep}              & RGB & 4413  & yes & no & no  \\
RENOIR      \cite{anaya2018renoir}          & RAW & 1500  & yes & no & no  \\
LOL         \cite{Chen2018Retinex}          & RGB & 500   & yes & no & no  \\
DeepUPE     \cite{xulearning}               & RGB & 3000  & no  & no & no  \\
VE-LOL-L    \cite{liu2021benchmarking}      & RGB & 2500  & yes & no & no  \\
DarkVision  \cite{zhang2023darkvision}      & RAW & 13455 & yes & no & no  \\ \hline
\multicolumn{1}{|l|}{Our} &
  \multicolumn{1}{l|}{RAW} &
  \multicolumn{1}{l|}{1500} &
  \multicolumn{1}{c|}{yes} &
  \multicolumn{1}{c|}{yes} &
  \multicolumn{1}{c|}{yes} \\ \hline
\end{tabular}%
}
\caption{Comparison with previous datasets.}
\label{tab:dataset_comp}
\end{table}

%% file: results_table.tex
\begin{table*}[t]
\caption{For all methods, the input exposure for both training and testing is 0.1s. $\Rightarrow$ denotes the ground-truth images used for training. The bold are the two best results. As can be seen, our model outperforms all other methods. See text for more details.}
\begin{center}
\begin{tabularx}{\textwidth}{|c| *{3}{Y|} }
\hline
 Train/Test & \multicolumn{2}{c|}{1s} & \multicolumn{2}{c|}{5s} & \multicolumn{2}{c|}{10s} \\
  &PSNR & SSIM & PSNR & SSIM & PSNR & SSIM \\
 \hline
\multicolumn{7}{c}{A - Single Exposure Baseline \label{results_output_A}} \\ \hline
\hline
SID\cite{chen2018learning}  $\Rightarrow$ 1  & {\bf \underline{38.17}} &{\bf \underline{ 0.95}} & 30.7 & 0.87 & 27.7 & 0.84 \\ 
SID\cite{chen2018learning}  $\Rightarrow$ 5  & 36.82 & 0.94 & {\bf \underline{33.35}} & {\bf \underline{0.91}} & 28 & 0.86 \\ 
SID\cite{chen2018learning} $\Rightarrow$ 10   & 34.88 & 0.9 & 30.52 & 0.88 & {\bf \underline{30}} & {\bf \underline{0.88}} \\
\hline
\hline
\multicolumn{7}{c}{B - Multi Exposure Baseline \label{results_output_B}} \\ \hline
\hline
SID\cite{chen2018learning} $\Rightarrow$ 1,5,10    & 35.77 & 0.92  &   29.55 &0.86 &26.25  & 0.82 \\
Retinex \cite{wei2018deep} $\Rightarrow$ 1,5,10   & 16.29 & 0.08  & 15.15 & 0.12 & 13.67  & 0.16 \\
\hline
\multicolumn{7}{c}{C - Two Exposure Interpolation\label{results_output_C}} \\ \hline
\hline
AdaFM\cite{he2019modulating}  $\Rightarrow$  1,10    & 37.86&0.85 &30.51 &0.73 &26.95  & 0.72 \\
CResMD\cite{he2020interactive} $\Rightarrow$ 1,10     & 36.37&0.8 &21.63 &0.46 &26.52  & 0.64 \\
Ours $\Rightarrow$ 1,10 & {\bf 38.17} & {\bf 0.95} &  {\bf 32.35} &  {\bf 0.89}&  {\bf29.67} & {\bf 0.87} \\ \hline

\multicolumn{7}{c}{D -  Two Exposure Extrapolation\label{results_output_D}} \\ \hline
\hline
AdaFM\cite{he2019modulating}  $\Rightarrow$ 1,5    & 37.86&0.85 &31.12 &0.76 &25.98  & 0.7 \\
CResMD\cite{he2020interactive}  $\Rightarrow$ 1,5  & 34.97&0.73 &23.73 &0.59 &16.17  & 0.17 \\
Ours  $\Rightarrow$ 1,5 & {\bf 38.17} &  {\bf 0.95} & {\bf 31.78} &  {\bf 0.89} & {\bf 28.65} &  {\bf 0.86} \\

\hline
\end{tabularx}
\end{center}

\label{results_output}
\end{table*}